# SUBCRITICAL STATES OF ASYMMETRIC NUCLEAR MATTER

A.I. Sanzhur[1*], S. Shlomo[2]

[1]*Institute for Nuclear Research, National Academy of Sciences of Ukraine, Kyiv, Ukraine*
[2]*Cyclotron Institute, Texas A&M University, College Station, TX 77840, USA*

*Corresponding author: andriy.sanzhur@gmail.com

The isobaric caloric curve is considered in subcritical states region. The energy fluctuations along the caloric curve are determined for small nuclear systems which consist of limited number of nucleons. The temperature dependence of heat capacity at fixed pressure is obtained. The calculated quantities of small nuclear system are discussed and checked against the ones for the infinite nuclear matter.

*Keywords:* caloric curve, energy fluctuation, small nuclear system

## 1. Introduction

Caloric curves measurements in heavy-ion collisions [1 – 3] have shown signs of a liquid-vapor phase transition in nuclear matter. The caloric curve, that is, the dependence of temperature, $T$, on excitation energy per particle $\varepsilon_{ex}$, provides important information on equation of state of nuclear matter. The plateau region of the caloric curve where one observes small change in temperature within the wide range of excitation energy gives the signal of phase transition [4]. This flat region of the caloric curve is accompanied by an increase of energy fluctuations and demonstrates departure from the Fermi-gas-like equation of state. For finite nuclear systems the extension of concepts as applied in the case of infinite matter is possible [5]. Statistical mechanics allows to find signatures of liquid-vapor phase transition for finite nuclear systems composed of a limited number of neutrons $N$ and protons $Z$ (small systems) [6]. Generally, the liquid-vapor phase transition is accompanied by an increase of energy fluctuations. In this context it is of interest to consider fluctuations for thermodynamic states along the caloric curve. Here we address this issue within the Isothermal-Isobaric Ensemble formulation. Calculations are performed at subcritical (i.e. below critical) values of temperature $T$ and pressure $P$.

In Section 2 the definitions of partition sum and Gibbs free energy are considered. The energy and enthalpy fluctuations are obtained. Section 3 is devoted to phase equilibrium in asymmetric nuclear matter. The relations for fractions of the total number of nucleons redistributed between liquid and vapor phases are determined. The shapes of the caloric curves and specific heat are obtained in Section 4 for small nuclear systems as well as for the infinite nuclear matter. Energy fluctuations for small systems are also calculated. Concluding remarks are summarized in Section 5.

## 2. Partition sum and Gibbs free energy. Energy fluctuation

In the macroscopic limit of very large volume and particle number (within the habitual thermodynamics for infinite nuclear matter) all the ensembles give the same description of energy as linear homogeneous function of entropy, volume and particle number. This is not the case for the small systems of a limited number of particles even though the surface, curvature and Coulomb effects are neglected. Nevertheless, for the certain set of environment variables, like pressure, $P$, and temperature, $T$, in our case, the thermodynamics of small system can be built [6]. In this paper we consider caloric curves at fixed value of pressure. In order to describe isobaric caloric curves, it is convenient to use Isothermal-Isobaric Ensemble formalism.

The partition sum $\Delta$ of Isothermal-Isobaric Ensemble is given by

$$\Delta(N,Z,P,T) = \sum_{i,V} \exp[(-PV - E_i)/T] = \sum_{V} \exp[(-PV - F(N,Z,V,T))/T] \,, \quad (1)$$

where $P$ is the pressure, $\{E_i\}$ is the energy spectrum of small nuclear system. In the right-hand side of Eq. (1) the summation (integration) is carried out over volume $V$, and $F = F(N,Z,V,T)$ stands for the free energy of nuclear system consisting of $N$ neutrons and $Z$ protons. It is assumed that the free energy $F$ can be scaled to given particle composition $A = N + Z$ from the Thomas – Fermi free energy per particle $\phi_{\text{TF}}(\rho_n, \rho_p, T)$ [4] as $F = A\phi_{\text{TF}}(\rho_n, \rho_p, T)$, where $\rho_n = N/V$ and $\rho_p = Z/V$ are, respectively, the neutron and proton densities. The thermodynamic potential of Isothermal-Isobaric Ensemble, that is Gibbs free energy $G$, is written as

$$G(N,Z,P,T) = -T\ln\Delta(N,Z,P,T) \,. \quad (2)$$

As seen from (2) the pressure and temperature are natural environment variables of Gibbs free energy. In order to account for situation that system can be found in two-phase thermodynamic state (liquid + vapor) the partition sum yields

$$\Delta(N,Z,P,T) = \sum_{\substack{N^{\text{liq}}+N^{\text{vap}}=N, \\ Z^{\text{liq}}+Z^{\text{vap}}=Z, \\ V^{\text{liq}}, V^{\text{vap}}}} \exp[(-P(V^{\text{liq}} + V^{\text{vap}}) - F^{\text{liq}} - F^{\text{vap}})/T] \,, \quad (3)$$

where $F^{liq} = F(N^{liq}, Z^{liq}, V^{liq}, T)$, $F^{vap} = F(N^{vap}, Z^{vap}, V^{vap}, T)$, and superscripts "liq" and "vap" denote liquid and vapor phases, respectively. The partition sum in Eq. (3) takes the account of particles redistributed between liquid and vapor phases provided the total number of neutrons and protons are fixed. The average value of energy, $\langle E \rangle$, and its dispersion (the energy fluctuation $\sigma_E$ squared), $\sigma_E^2 = \langle E^2 \rangle - \langle E \rangle^2$, are obtained from the Gibbs free energy $G$ as

$$\langle E \rangle = G - T \left(\frac{\partial G}{\partial T}\right)_{N,Z,P} - P \left(\frac{\partial G}{\partial P}\right)_{N,Z,T}, \tag{4}$$

and

$$\langle E^2 \rangle - \langle E \rangle^2 = -T^3 \left(\frac{\partial^2 G}{\partial T^2}\right)_{N,Z,P} - 2PT^2 \left(\frac{\partial^2 G}{\partial P \partial T}\right)_{N,Z} - P^2 T \left(\frac{\partial^2 G}{\partial P^2}\right)_{N,Z,T}. \tag{5}$$

The excitation energy per particle $\varepsilon_{ex}$, needed for the determination of the caloric curve, $T(\varepsilon_{ex})$, is obtained from Eq. (4) as

$$\varepsilon_{ex} = (\langle E \rangle - E_{gs})/A, \tag{6}$$

where $E_{gs}$ is the ground state energy at $T = 0$ and fixed value of $P$.

Combined with the energy fluctuation it is worthwhile to consider also the fluctuation of the enthalpy $\langle H \rangle = \langle E \rangle + P \langle V \rangle = G - T(\partial G/\partial T)_{N,Z,P}$ since the enthalpy is natural heat function for variables $P$ and $T$. The dispersion of the enthalpy, $\sigma_H^2 = \langle H^2 \rangle - \langle H \rangle^2$, is directly concerned with the heat capacity at fixed pressure, $C_P$, as follows [7]

$$\langle H^2 \rangle - \langle H \rangle^2 = T^2 \left(\frac{\partial \langle H \rangle}{\partial T}\right)_{N,Z,P} = C_P T^2. \tag{7}$$

The fluctuation $\sigma_H$ and heat capacity $C_p$ become very large in a two-phase region at constant pressure and temperature due to the fact that heat can be absorbed converting some of one phase into the other. So, strong increase in heat capacity is an additional sigh that phase transition occurs.

### 3. Infinite matter

To highlight the difference of a small nuclear system against infinite matter one needs to consider the macroscopic limit of large nucleon number. Let take a look at nuclear system having so large particle number $A = N + Z$ that make possible the use of regular thermodynamics. The Gibbs free energy per particle $g(P, T, X) = G/A = (\partial G/\partial A)_{P,T,X}$, where $X = (N - Z)/A$ is the asymmetry parameter, is a function of intensive variables only. Introducing isovector, $\mu_0 = g - X(\partial g/\partial X)_{P,T}$, and isoscalar, $\mu_1 = (\partial g/\partial X)_{P,T}$, chemical potentials, the Gibbs conditions for liquid-vapor equilibrium [8] are written as

$$\mu_0^{\text{liq}}(P,T,X^{\text{liq}}) - \mu_0^{\text{vap}}(P,T,X^{\text{vap}}) = 0, \qquad (8)$$

$$\mu_1^{\text{liq}}(P,T,X^{\text{liq}}) - \mu_1^{\text{vap}}(P,T,X^{\text{vap}}) = 0. \qquad (9)$$

We note, that within the macroscopic limit there is equivalence of Isothermal-Isobaric and Canonical Ensemble. The pressure in Eqs. (8), and, (9), which should be the same for liquid and vapor phases in equilibrium, is associated with the equation of state $P = P(\rho, X, T)$ for infinite nuclear matter [9], where $\rho = A/V$ is the nucleon density. With regard to the mentioned equation of state one has two solutions of Eqs. (8), and, (9), that are $\{\rho^{\text{liq}}, X^{\text{liq}}\}$ and $\{\rho^{\text{vap}}, X^{\text{vap}}\}$, at the certain value of temperature $T$.

For two-phase part of the isobar ($P = const$) the energy per particle, $\varepsilon = E/A$, is determined as sum of contributions from each phase,

$$\varepsilon = \alpha^{\text{liq}} \varepsilon^{\text{liq}} + \alpha^{\text{vap}} \varepsilon^{\text{vap}}, \qquad (10)$$

where $\alpha^{\text{liq}} = A^{\text{liq}}/A$ and $\alpha^{\text{vap}} = A^{\text{vap}}/A$ are the fractions of the total number of particles $A$, redistributed between liquid and vapor phases, respectively. From the conservation conditions for the total number of particles and neutron excess of the whole system determined by the particle $A$ and asymmetry parameter $X$, namely,

$$\alpha^{\text{liq}} + \alpha^{\text{vap}} = 1, \quad \alpha^{\text{liq}} X^{\text{liq}} + \alpha^{\text{vap}} X^{\text{vap}} = X, \qquad (11)$$

one obtains the values of particle fractions:

$$\alpha^{\text{liq}} = (X^{\text{vap}} - X)/(X^{\text{vap}} - X^{\text{liq}}), \quad \alpha^{\text{vap}} = (X - X^{\text{liq}})/(X^{\text{vap}} - X^{\text{liq}}). \qquad (12)$$

By virtue of equilibrium conditions (8), and, (9), the above particle fractions are, in general, $P$- and $T$-dependent. In case of a single phase one has to put $\alpha^{\text{liq}} = 1$, $\alpha^{\text{vap}} = 0$, $X^{\text{liq}} = X$ for the liquid or $\alpha^{\text{liq}} = 0$, $\alpha^{\text{vap}} = 1$, $X^{\text{vap}} = X$ for the vapor phase. The construction of the caloric curve requires the calculation of the excitation energy per particle $\varepsilon_{\text{ex}}$. The value of $\varepsilon_{\text{ex}}(T,P,X)$ is obtained by subtracting the corresponding ground state value of energy per particle $\varepsilon(T=0,P,X)$ from the corresponding value (10) at certain value of $T > 0$:

$$\varepsilon_{\text{ex}}(T,P,X) = \varepsilon(T,P,X) - \varepsilon(T=0,P,X). \qquad (13)$$

Similarly to the energy per particle of the asymmetric nuclear matter we determine the value of enthalpy per particle $h = H/A = \varepsilon + P/\rho$, where $H = E + PV$ is the enthalpy. Just as in Eq. (10), the enthalpy per particle for the two-phase equilibrium state is given by

$$h = \alpha^{\text{liq}} h^{\text{liq}} + \alpha^{\text{vap}} h^{\text{vap}}. \qquad (14)$$

Relying on Eq. (14) one can obtain the specific heat per particle at fixed pressure, $c_P = C_P/A$, which is defined by

$$c_P = \left(\frac{\partial h}{\partial T}\right)_{P,X}. \tag{15}$$

In contrast to the case of a single phase, the enthalpy per particle in the form of Eq. (14) contains particle fractions $\alpha^{\text{liq, vap}}$ which themselves are temperature dependent. This usually leads to increase in the value of the specific heat as compared to the corresponding single-phase value. In order to obtain the temperature dependence of particle fractions $\alpha^{\text{liq, vap}}(T)$ one may differentiate equilibrium conditions (8), and, (9), with respect to temperature assuming that their solutions for $X^{\text{liq, vap}}$ depend on $T$. As a result one arrives at two linear equations with regard to the derivatives $(\partial X^{\text{liq, vap}}/\partial T)_{P,X}$. Then, resolving the obtained equations, the temperature derivatives of particle fractions $(\partial \alpha^{\text{liq, vap}}/\partial T)_{P,X}$ can be found straightforwardly using relations (12).

## 4. Results and discussion.

We study the shape of caloric curve and energy fluctuations along this curve for the case of isobaric heating. For this purpose we follow the temperature dependent Thomas – Fermi theory using the Skyrme-type force as the effective nucleon-nucleon interaction [4, 10], see also Appendix A. We have calculated the isobaric caloric curves, using $P = 0.05$ MeV/fm$^3$, for several small nuclear systems having different particle numbers $A = 48, 120$, and $216$ at the same asymmetry parameter $X = 1/6$. By the neutron-proton composition these systems correspond to $^{48}$Ca, $^{120}$Sn, and $^{216}$Th nuclei. Calculation was carried out for the temperature interval $T = 5 \div 12$ MeV using KDE0v1 Skyrme nucleon-nucleon effective interaction [11]. The value of pressure and the temperature interval correspond to subcritical thermodynamic states of nuclear matter. Recall that for the Skyrme force KDE0v1 the values of pressure and temperature at critical point are about of $P_{\text{cr}} = 0.22$ MeV/fm$^3$ and $T_{\text{cr}} = 14.8$ MeV. The results for $A = 48, 120$, and $216$ are shown in Fig. 1 by the dash-dotted, dashed, and dotted lines, respectively (see Eqs. (4), and, (6)). For the purpose of comparison, the calculation at the same pressure and asymmetry parameter was carried out for infinite asymmetric nuclear matter (solid line in Fig. 1, see Eqs. (10), and, (13)). Comparing the dot-dashed, dashed and dotted lines, with the solid one in Fig. 1, it is seen that the temperature in the middle of plateau region for the small systems is lower than that for infinite matter. Also, the results obtained for small systems are smooth and do not demonstrate fracture (derivative discontinuity) which is seen for infinite matter at bubble and dew points shown in Fig. 1 by arrows.

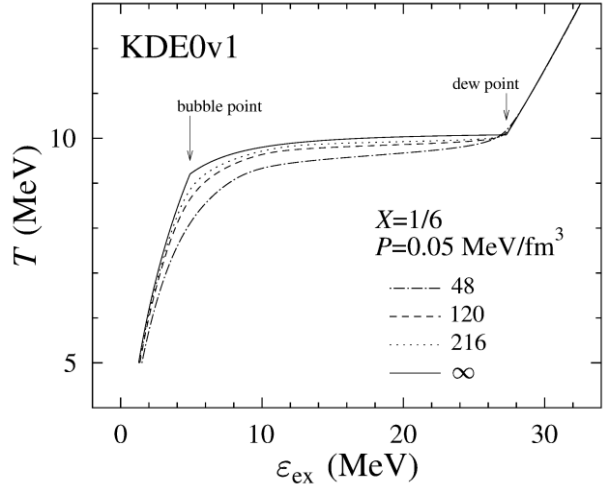

Fig. 1. Isobaric caloric curves $T(\varepsilon_{ex})$ obtained at pressure $P = 0.05\,\text{MeV/fm}^3$ and neutron-proton asymmetry parameter $X = 1/6$. Dot-dashed, dashed, and dotted lines present the results for small nuclear systems with $A = 48, 120$, and 216, respectively. Solid line gives the result in the case of infinite matter. The positions of bubble and dew points of infinite matter are shown by arrows. Calculations were carried out using KDE0v1 Skyrme nucleon-nucleon interaction [11].

We also obtained the energy dispersion $\sigma_E^2 = \langle E^2 \rangle - \langle E \rangle^2$ by means of Eq. (5). As seen from Eq. (5) the calculation of the dispersion and, consequently, the fluctuation of energy requires the values of second derivatives of the Gibbs thermodynamic potential $G$ with respect to the temperature and pressure. Figure 2 presents the energy dispersions for small nuclear systems with $A = 48, 120$, and 216 as functions of excitation energy per particle along the corresponding caloric curves shown in Fig. 1. Figure 2 demonstrates the increase of energy dispersions in the two-phase region of excitation energies between bubble and dew points. Such an increase, together with the plateau region in caloric curve $T(\varepsilon_{ex})$, gives the signature of the occurring phase transition. The presented results for small nuclear systems could be valuable to give an idea on the excitation energy range where to expect the observation of liquid-vapor phase transition.

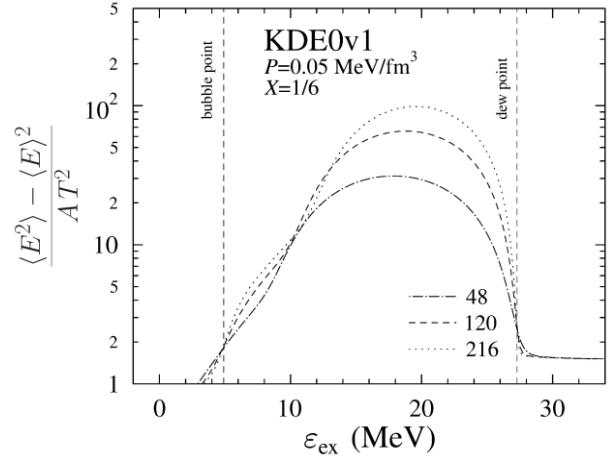

Fig. 2. The energy dispersion per particle over square of temperature versus the excitation energy per nucleon $\varepsilon_{\rm ex}$, see Eqs. (5), and (6). Results are obtained for small nuclear systems with $A = 48$ (dot-dashed line), 120 (dashed line), and 216 (dotted line) along the corresponding caloric curves, see Fig. 1. The range of $\varepsilon_{\rm ex}$ between bubble and dew points (vertical dashed lines) corresponds to coexistence of liquid and vapor phases for the case of infinite nuclear matter.

Caloric curves illustrated in Fig. 1 are almost flat in the phase coexistence region where they are located within the small temperature interval of about 1 MeV wide. It is of interest to see the behavior of the heat capacity for this quite narrow temperature region. The temperature dependence of the heat capacity per particle at fixed pressure, $c_P$, is displayed in Fig. 3. Calculations were performed using Skyrme nucleon-nucleon effective interaction KDE0v1 at the same values of pressure and asymmetry parameter as for Figs. 1 and 2. The result of heat capacity calculation for infinite nuclear matter, see Eq. (15), is presented in Fig. 3 by the solid line. The value of $c_P$ exhibits two abrupt jumps, the one on the left-hand side of the figure correspond to the temperature of the bubble point, and the other one on the right which correspond to the dew point temperature. The behavior of $c_P$ between the bubble and dew point temperatures demonstrates the strong increase in the value of heat capacity. Based on Eq. (7), the values of $c_P$ were obtained for small nuclear systems having numbers of nucleons $A = 48$ (dot-dashed line in Fig. 3), 120 (dashed line), and 216 (dotted line). As seen from Fig. 3, in contrast to the case of infinite matter, the heat capacity of a small system consisting of limited number of nucleons is a smooth function of temperature without abrupt jumps. It is peaked at certain temperature in the considered range of phase coexistence with very high maximum value (as compared to the ideal gas value of 5/2). The peak temperature and the maximum value increase as the nucleon number of the system goes higher, compare dot-dashed, dashed and dotted lines in Fig. 3. In a small nuclear system, considering the phase equilibrium of dense particle aggregates (clusters) with the surrounding saturated vapor of nucleons, the existence of strong peak in the specific heat $c_P$ at subcritical temperature is an evidence of the liquid-vapor phase transition.

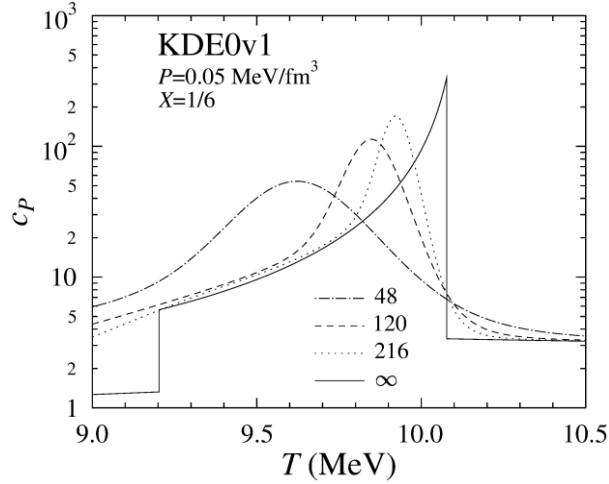

Fig. 3. Heat capacity per particle at fixed pressure as a function of temperature. Presented results correspond to the values of pressure $P = 0.05$ MeV/fm³ and asymmetry parameter $X = 1/6$. Solid line shows the result for infinite nuclear matter. Dot-dashed, dashed and dotted lines are obtained for small nuclear systems having particle numbers $A = 48$, 120, and 216, respectively.

## 4. Summary

Caloric curves $T(\varepsilon_{\text{ex}})$ for small nuclear systems of limited number of nucleons was considered in comparison with the infinite nuclear matter. The study of small (finite) systems requires the application of statistical mechanics, whereas to describe the infinite matter it is sufficient to apply equilibrium conditions from regular thermodynamics. Calculations of the caloric curves $T(\varepsilon_{\text{ex}})$, energy dispersion $\sigma_E^2(\varepsilon_{\text{ex}})$, and heat capacity per particle $c_P(T)$, were carried out using the KDE0v1 Skyrme-type nucleon-nucleon interaction. As evident from the calculations for infinite matter, see solid lines in Figs. 1 and 3, the transfer between the single phase state and the state of phase coexistence (bubble and dew points) is accompanied by irregularities, like fracture in $T(\varepsilon_{\text{ex}})$ or abrupt jump in $c_P(T)$. As for the finite system, all the obtained quantities demonstrate smooth behavior. The flat region of the caloric curve, high maximum in heat capacity, and high value of energy fluctuation give us signs of the occurring liquid-vapor phase transition in the finite system. The presented results on energy fluctuations for small nuclear systems could be valuable to give an idea on the excitation energy range where to expect the observation of liquid-vapor phase transition.

This work is partially supported by the project №0120U102221 of the National Academy of Sciences of Ukraine (A.I. Sanzhur), and partially by the US Department of Energy under grant DOE-FG03-93ER40733 (S. Shlomo).

**Appendix A. Skyrme nucleon-nucleon effective interaction**

In this paper we adopt the following form of the Skyrme nucleon-nucleon effective interaction $U_{12}(\mathbf{r}_1, \mathbf{r}_2; \rho)$ [12, 13]:

$$U_{12}(\mathbf{r}_1, \mathbf{r}_2; \rho) = t_0(1 + x_0 P_{12}^\sigma)\delta(\mathbf{r}_1 - \mathbf{r}_2)$$

$$+ \frac{1}{2} t_1 (1 + x_1 P_{12}^\sigma) \times \left[ \overleftarrow{k}_{12}^2 \delta(\mathbf{r}_1 - \mathbf{r}_2) + \delta(\mathbf{r}_1 - \mathbf{r}_2) \vec{k}_{12}^2 \right]$$

$$+ t_2(1 + x_2 P_{12}^\sigma) \overleftarrow{k}_{12} \delta(\mathbf{r}_1 - \mathbf{r}_2) \vec{k}_{12} \quad (A1)$$

$$+ \frac{1}{6} t_3 (1 + x_3 P_{12}^\sigma) \rho^\nu \left( \frac{\mathbf{r}_1 + \mathbf{r}_2}{2} \right) \delta(\mathbf{r}_1 - \mathbf{r}_2)$$

$$+ i W_0 \overleftarrow{k}_{12} \delta(\mathbf{r}_1 - \mathbf{r}_2)(\boldsymbol{\sigma}_1 + \boldsymbol{\sigma}_2) \times \vec{k}_{12},$$

where $t_i$, $x_i$, $\nu$, and $W_0$ are parameters of the interaction and $P_{12}^\sigma$ is the spin exchange operator, $\boldsymbol{\sigma}_i$ is the Pauli spin operator, $\vec{k}_{12} = -i(\vec{\nabla}_1 - \vec{\nabla}_2)/2$, $\overleftarrow{k}_{12} = -i(\overleftarrow{\nabla}_1 - \overleftarrow{\nabla}_2)/2$. Here, the right and left arrows indicate momentum operators act on the right (initial state) and on the left (final state), respectively. The nucleon-nucleon interaction (A1) includes central, non-local (momentum-dependent), density dependent, and spin-orbit terms. The set of parameters for the KDE0v1 Skyrme-type nucleon-nucleon interaction is presented in the Table below.

*Table*

| Parameter | Value (KDE0v1) |
|---|---|
| $t_0$ (MeV·fm$^3$) | $-2553.0843$ |
| $t_1$ (MeV·fm$^5$) | $411.6963$ |
| $t_2$ (MeV·fm$^5$) | $-419.8712$ |
| $t_3$ (MeV·fm$^{3(1+\nu)}$) | $14603.6069$ |
| $x_0$ | $0.6483$ |
| $x_1$ | $-0.3472$ |
| $x_2$ | $-0.9268$ |
| $x_3$ | $0.9475$ |
| $\nu$ | $0.1673$ |
| $W_0$ (MeV·fm$^5$) | $124.4100$ |

*Note.* Parameters of Skyrme nucleon-nucleon interaction KDE0v1 [11].

The use of interaction (A1) within the semiclassical Thomas – Fermi theory [10] allows to express Skyrme–Hartree–Fock energy as a functional of the local neutron and proton densities and build energy-density functional eliminating the single-particle wave functions. The temperature dependent Thomas – Fermi approximation, as applied to infinite nuclear matter of uniform particle density, provides the expression for the free energy per particle $\phi_{TF}$, namely,

$$\rho\phi_{TF}(\rho,X,T) = T\sum_q \left(\eta_q\rho_q - \frac{2}{3}\mathcal{A}_q^* J_{3/2}(\eta_q)\right)$$
$$+ \frac{1}{2}t_0\left[\left(1+\frac{x_0}{2}\right)\rho^2 - \left(x_0+\frac{1}{2}\right)(\rho_n^2+\rho_p^2)\right] \quad (A2)$$
$$+ \frac{1}{12}t_3\rho^\nu\left[\left(1+\frac{x_3}{2}\right)\rho^2 - \left(x_3+\frac{1}{2}\right)(\rho_n^2+\rho_p^2)\right],$$

where $T$ is the temperature, $q$ is the isospin index, $\rho_q$ stand for the densities of particle species $q$ ($q = n$ for neutron and $q = p$ for proton), $\rho = \rho_n + \rho_p$ is the total density of nucleons, and $X = (\rho_n - \rho_p)/\rho$ is the asymmetry parameter. The Fermi integral $J_\beta(\eta_q) = \int_0^\infty dz\, z^\beta/(1+\exp(z-\eta_q))$ in Eq. (A2) depends on the fugacity $\eta_q$. The value of $\eta_q$ is obtained from the condition

$$\rho_q = \mathcal{A}_q^* J_{1/2}(\eta_q). \quad (A3)$$

Here, $\mathcal{A}_q^* = \frac{1}{2\pi^2}\left(\frac{2m_q^*T}{\hbar^2}\right)^{3/2}$ and $m_q^*$ is the effective nucleon mass derived by

$$\frac{\hbar^2}{2m_q^*} = \frac{\hbar^2}{2m} + \frac{1}{4}\left[t_1\left(1+\frac{x_1}{2}\right) + t_2\left(1+\frac{x_2}{2}\right)\right]\rho$$
$$+ \frac{1}{4}\left[t_2\left(x_2+\frac{1}{2}\right) - t_1\left(x_1+\frac{1}{2}\right)\right]\rho_q, \quad (A4)$$

where $m$ is the bare nucleon mass. The pressure in the two-phase equilibrium conditions (8), and, (9), of Section 3, is associated with the equation of state $P(\rho,X,T)$ which can be obtained from free energy per particle $\phi_{TF}(\rho,X,T)$ of Eq. (A2) as

$$P(\rho,X,T) = \rho^2\left(\frac{\partial\phi_{TF}}{\partial\rho}\right)_{X,T}. \quad (A5)$$

# REFERENCES

1. J. Cibor et al. Dynamic evolution and the caloric curve for medium mass nuclei. Phys. Lett. B 473 (2000) 29.
2. J.B. Natowitz et al. Limiting temperatures and the equation of state of nuclear matter. Phys. Rev. Lett. 89 (2002) 212701.
3. J.B. Natowitz et al. Caloric curves and critical behavior in nuclei. Phys. Rev. C 65 (2002) 034618.
4. V.M. Kolomietz, S. Shlomo. *Mean field theory* (Singapore: World Scientific, 2020) 565 p.
5. Tapas Sil et al. Liquid-gas phase transition in infinite and finite nuclear systems. Phys. Rev. C 69 (2004) 014602.
6. Terrell L. Hill. *Thermodynamics of Small Systems* (New York: Dover Publications, 2013) 416 p.
7. Terrell L. Hill. *Statistical Mechanics. Principles and Selected Applications* (New York, Dover Publications, 1987) 432 p.
8. L.D. Landau, E.M. Lifshitz. *Statistical physics. Part 1* (Oxford: Pergamon Press, 1980) 544 p.
9. V.M. Kolomietz et al. Equation of state and phase transitions in asymmetric nuclear matter. Phys. Rev C 64 (2001) 024315.
10. M. Brack, C. Guet, H.-B. Håkansson. Selfconsistent semiclassical description of average nuclear properties– a link between microscopic and macroscopic models. Phys. Rep. 123 (1985) 275.
11. B.K. Agrawal, S. Shlomo, V. Kim Au. Determination of the parameters of a Skyrme type effective interaction using the simulated annealing approach. Rhys. Rev. C 72 (2005) 014310.
12. D. Vautherin, D.M. Brink. Hartree-Fock Calculations with Skyrme's Interaction. I. Spherical Nuclei. Rhys. Rev C 5 (1972) 626.
13. E. Chabanat et al. A Skyrme parametrization from subnuclear to neutron star densities. Nucl. Phys. A 627 (1997) 710.